\documentclass[preprintnumbers]{revtex4}
\usepackage{graphicx}
\usepackage{color}
\usepackage{eurosym}
\usepackage{amssymb}
\usepackage{amsmath}
\usepackage{amsfonts}

\begin{document}

\title{Benchmark for a quasi-bound state of the ${K^-}pp$ system}
\author{Roman Ya. Kezerashvili$^{1,2}$, Shalva M. Tsiklauri$^{3}$, Igor N.
Filikhin$^{4}$, Vladimir M. Suslov$^{4}$, and Branislav Vlahovic$^{4}$}
\affiliation{\mbox{$^{1}$Physics Department, New
York City College of Technology, The
City University of New York,} \\
Brooklyn, NY 11201, USA \\
\mbox{$^{2}$The Graduate School and University Center, The
City University of New York,} \\
New York, NY 10016, USA \\
\mbox{$^{3}$Borough of Manhattan Community College, The City University of
New York,}\\
New York, NY 10007, USA \\
\mbox{$^{4}$North Carolina Central University, Durham, NC 27707, USA }}

\begin{abstract}
We present three-body nonrelativistic calculations within the framework of a
potential model for the kaonic cluster ${K^-}pp$ using two
completely different methods: the method of hyperspherical harmonics in the
momentum representation and the method of Faddeev equations in configuration
space. To perform a numerical benchmark, different $NN$ and antikaon-nucleon
interactions are applied. The results of the calculations for the ground
state energy for the ${K^-}pp$ system obtained by both methods are
in reasonable agreement. Although the ground state energy is not sensitive
to the $NN$ interaction, it shows very strong dependence on the ${\bar
K}N$ potential. We show that the dominant clustering of the ${K^-}
pp$ \ system in the configuration $\Lambda (1405)+p$ allows us to calculate
the binding energy to good accuracy within a simple cluster approach for the
differential Faddeev equations. The theoretical discrepancies in the binding
energy and width for the ${K^-}pp$ system related to the different 
$NN$ and ${\bar K}N$ interactions are addressed.
\end{abstract}

\maketitle

\section{Introduction}

\bigskip The theoretical and experimental studies of composite systems of
mesons and baryons is a challenging issue in hadronic and nuclear physics.
The study of the $\bar{K}N$ interaction and the formation of the $%
\Lambda (1405)$ as a quasi-bound state of $\bar{K}N$ \cite{Dalitz, Dalitz2} started in the 60s. Early suggestions for kaon condensate in
dense matter \cite{Kaplan, Brown, Brown1, Lee1996} motivated the search for bound
states of kaons in nuclei, since the kaon--nucleus interaction could answer
the question of whether kaon condensation takes place in the core of neutron
stars. Kaonic systems have attracted much attention in the last decade. In
Ref. \cite{Kishimoto} it was shown that kaonic nuclei could be produced by
the (${K^-},$ $p$) and (${K^-},$ $n$) reactions.
However, the renaissance in this field started after the prediction of
deeply bound $\bar{K}-$ states by calculations performed by Akaishi
and Yamazaki in Ref. \cite{AYPRC2002} and followed by several subsequent
publications \cite{AYPL2002, AYPL2004, ADYPL2005, DotePRC2004, AYKN}. \ The experimental search was
initiated at the KEK 12 GeV proton synchrotron in Japan by the finding of
peaks in the nucleon spectra of ${K^-}$-absorption in $^{4}$He 
\cite{Suzuki, Iwasaki}.

The $\bar{K}N$ interaction has been found to be strongly attractive
in the isospin $I=0$ channel and, as a consequence of that, nucleons are
attracted by the $\bar{K}$ meson to form dense nuclear states. In
1963, in Ref. \cite{Nogami1963} Nogami first examined three possible isospin
configurations of the $\bar{K}NN$ system and discussed the possible
existence of the bound state using a rather crude calculation. The primary
focus was on the lightest kaonic system ${K}^-pp$ because it has
two ${K}^-p$ pairs with isospin $I=0$ that can be considered as a
building block of a three-body kaonic cluster. Regardless of the fact that
two protons are not bound, there is an expectation that $\bar{K}N$
interactions can be strong enough to bound the ${K}^-pp$ system.
Interestingly enough, even two protons can be bound by a single $\bar%
{K}$ meson and it is assumed that this would be the strongest binding
cluster among the three-body systems!

On the experimental side, several experiments have been performed to search
for the kaonic ${K^-}pp$ cluster using various reactions. The
first measurement reported by the FINUDA collaboration was the measurement
of the invariant mass distribution of the $p$ produced by ${K^-}-$%
absorption by $^{6}$Li, $^{7}$Li, and $^{12}$C at the e$^{+}$e$^{-}$ rings
of the DA$\mathit{\Phi }$NE machine in Frascati \cite{AgnelloKpp}. The
analysis of the invariant mass distribution of the $\Lambda p$ produced by ${K^-} -$ absorption gave the value (2255$\pm $9) MeV/c$^{2}$ for
the mass of the peak, corresponding to the binding energy 115$_{-5}^{+6}($%
stat)$_{-4}^{+3}($syst) MeV and the width 67$_{-11}^{+14}($stat)$_{-3}^{+2}($%
syst) MeV. However, the authors of Refs. \cite{Oset1}, \cite{MagasOset}
claimed that the FINUDA data can be explained without postulating the
existence of the ${K^-}pp$ bound state and suggested that the
peaks of the proton spectra come from the ${K^-}$ absorption on a
pair of nucleons, leaving the rest of the nucleons as spectators.
Interestingly enough, even though such a mechanism cannot be completely
excluded, there exists the further counter-arguments \cite{CounterYamazaki}
against the interpretations of stopped-${K^-}$ experimental data
of the FINUDA group by \cite{MagasOset}. The OBELIX experiment at LEAR-CERN 
\cite{Bendiscioli} also suggested the observation of the ${K^-}pp$
state with invariant-mass spectroscopy of stopped $\overset{\_}{p}$
reactions on $^{4}$He. However, a less significant signal was reported in
the stopped-$\overset{\_}{p}$ reaction. After the FINUDA collaboration
observation \cite{AgnelloKpp}, there have been reports of possible
experimental evidence for the kaonic ${K^-}pp$ state in heavy ion
collisions. Results were obtained by the FOPI group experiment \cite{FOPI}
at GSI in the study of Ni$+$Ni and Al$+$Al collisions with the production of 
$\Lambda p$ and $\Lambda d$. The DISTO collaboration at the SATURNE machine
have reanalyzed their dataset of the experiment on the exclusive $%
pp\rightarrow p\Lambda K^{+}$ reaction at 2.85 GeV to search for a
strongly-bound compact ${K^-}pp$ state formed from the $%
p+p\rightarrow K^{+}+({K^-}pp)$ reaction \cite{DISTO1}, \cite%
{DISTO2}. The enormously large bump structure observed in the $K^{+}$
missing mass and the $p\Lambda $ invariant-mass spectra at 2.85 GeV indicate
a possible candidate for the formation of a compact ${K^-}pp$
cluster with large binding energy. The binding energy and width were
determined to be 103$\pm $3(stat)$\pm $5(syst) MeV and 118$\pm $8(stat)$\pm $%
10(syst) MeV, respectively. Therefore, two collaborations \cite{AgnelloKpp}
and \cite{DISTO1}, \cite{DISTO2} claim that ${K^-}pp$ is strongly
bound with a binding energy of more than 100 MeV, although the decay width
is rather different between these two. The first search results using the $%
\gamma d\rightarrow K^{+}\pi ^{-}X$ reaction at the range of photon energy
1.5-2.4 GeV were reported in Ref. \cite{LEPSColab}. A statistically
significant peak structure was not observed in the differential cross
section of the $K^{+}\pi ^{-}$ photo-production in the region from 2.22 to
2.36 GeV/c$^{2}$ in the inclusive missing mass spectrum. However, the upper
limits of the differential cross section of the ${K^-}pp$ bound
state production were determined with the assumed widths of 20~MeV, 60 MeV
and 100 MeV.

Ref. \cite{JPARCE15} reported on preliminary results of the J-PARC E15
experiment aiming to search for the bound state of ${K^-}pp$ via
the in-flight $^{3}$He$(K^{-},n)$ reaction at 1 GeV/c and the first data
collection was performed with 5$\times $10$^{9}$ incident kaons on the $^{3}$%
He target. The experiment investigates the ${K^-}pp$ bound state
exclusively both in the formation via missing-mass spectroscopy and its
decay via invariant-mass spectroscopy using the emitted neutron and the
expected decay, ${K^-}pp\rightarrow \Lambda $ $pp\rightarrow \pi
^{-}pp$, respectively. Data analyses of the semi-inclusive and exclusive
channels are in progress. At the same time, in Ref. \cite{JPARCE15N} have
been reported results for the in-flight kaon-induced reaction on $^{3}$He: ${K^-}$ +$^{3}$He$\rightarrow {K^-}pp+n$. In particular,
the semi-inclusive analysis of the neutron missing-mass spectrum, no
significant peak was observed in the region corresponding to ${K^-}%
pp$ binding energy larger than 80 MeV, where a bump structure was reported
in the $\Lambda p$ final state in different reactions by FINUDA and DISTO
collaborations. Thus, the situation is still controversial and the existence
of the ${K^-}pp$ bound state has not been established yet. New
experiments using different reactions could help to resolve this
controversial situation.

Such experiments are being planned and performed by the HADES collaboration 
\cite{HADES}. It was recently proposed to search for kaonic nuclei at the
future SuperB factory, which will be built in the Tor Vergata University
Campus in Rome. The ${K^-}pp$ cluster can be identified through
its $\Lambda p$ decay mode, searching for a narrow peak at a mass of about
2.25 GeV/c$^{2}$ in the $\Lambda p$ invariant mass spectrum and measuring
the $\Lambda -p$ angular correlations, which can give important hints on the
nature of the event \cite{SuperB}. One of the advantages of searching for
light ${K^-}-$nuclear clusters at the future SuperB factory is
that the search could be extended from the nuclear medium to the vacuum,
looking for its production in the strong decays of $\Upsilon $(1S) by taking
advantage of the high luminosity of these machines. Also, all problems
related to the influence of the medium as well as the final state
interaction will hopefully be avoided.

On the theoretical side, there are many reports related to $\bar{K}$ 
$-$nuclear systems. Today, refinements of the existing methods for studying
few-body systems, on one hand, and developments of new methods, on the other
hand, and advances in computational facilities enable very precise
calculations for few-body $\bar{K}-$nuclear systems. The light
kaonic $\bar{K}NN$ cluster represents a three-body system and has
been treated in the framework of various theoretical approaches such as
variational methods \cite{AYPL2002, AYPL2004, ADYPL2005, DotePRC2004, AYKN, DoteProg2002, DW2007, DHWnp2008, DHW, WycechGreen, ChinesePhys, FeshbaxRes2014}, 
the method of Faddeev equations \cite{Nina1, Nina2, IkedaSato1, IkedaSato2, Oset, IkedaKamanoSato3, FaddFixCent1, FaddFixCent2, 
FaddFixCent3, FaddFixCent4, FadYa4body, Nina2014} and the method of hyperspherical harmonics (HH) \cite{BarneaK, RKSh.Ts2014}.

In the framework of the Brueckner-Hartree-Fock theory, by defining the $%
\bar{K}N$\ $\ g$-matrix in a nuclear medium as well as using
antisymmetrized molecular dynamics for the spatial functions, the ${K}^-pp$ cluster has been studied in Refs. \cite{AYPL2002}-\cite{AYPL2004}, 
\cite{ChinesePhys}. Variational calculations of the ${K}^-pp$
system using several versions of energy-dependent effective $\bar{K}%
N $ interactions derived from chiral SU(3) dynamics were performed in Refs. 
\cite{DHWnp2008}, \cite{DHW}. A different ansatz was used in Ref. \cite%
{WycechGreen} for the trial wave function of strongly correlated nucleons in
the variational calculations of the binding energy and width for the $%
{K}^-pp$ cluster. Recently in Ref. \cite{FeshbaxRes2014} the $%
{K}^-pp$ cluster was investigated by combining a coupled-channel
Complex Scaling Method with the Feshbach resonance method allowing one to
effectively treat a coupled-channel problem as a single channel problem. The
authors use an energy-dependent chiral-theory based potential and employ the
correlated Gaussian function \cite{VargaStochastic}. They report that the $%
{K}^-pp$ cluster is shallowly bound with a binding energy of
around 20-35 MeV and the half decay width ranges from 20 to 65 MeV.

The strongly bound nature of the ${K}^-pp$ system is shown in
Refs. \cite{Nina1, Nina2, Nina2014} by using the
coupled-channel Faddeev equations in the Alt-Grassberger-Sandhas form \cite%
{AGS} and solving for all two-body separable potentials to find the
resonance energies and widths. Similar calculations have been done in Refs. 
\cite{IkedaSato1, IkedaSato2} and \cite{IkedaKamanoSato3}. The $%
{K}^-pp$ cluster is studied in Refs. \cite{FaddFixCent1, FaddFixCent2, FaddFixCent3, FaddFixCent4} using the fixed center approximation to the Faddeev equations
that relies upon on shell two body amplitudes. Thus, the structure and the
production mechanism of the ${K}^-pp$ bound state have been
investigated using various theoretical approaches. All the aforementioned
approaches predict the existence of a bound state for the $\bar{K}NN$
system. While on the one hand, various theoretical approaches predict a
strong bound state of the ${K}^-pp$ system, on the other hand,
various theories based on chiral dynamics predict a shallow bound state for
the ${K}^-pp$ cluster. The predicted values for the binding energy
and the width are in considerable disagreement: 9--95 MeV and 20--110 MeV,
respectively. The serious theoretical discrepancies concerning ${K}^-pp$ 
binding and width essentially come from different starting ansatzes:
phenomenological\emph{\ }$\bar{K}$-nucleon potential constructed
using $\bar{K}N$ scattering data and the data of kaonic hydrogen
atomic shift together with the ansatz that $\Lambda $(1405) resonance is a
bound state of $\bar{K}N$ system, a chiral ansatz for $\bar{K%
}N$ interaction that leads to a substantially shallow $\bar{K}N$
potential, and the calculation methods. Further theoretical investigations
are apparently needed to resolve this controversial situation that perhaps
stems from the ambiguity of the $\bar{K}N$ interaction, the
importance of $NN$ interaction or the different procedures for three-body
calculations.

We present three-body nonrelativistic calculations within the framework of a
potential model for the kaonic cluster ${K}^-pp$ applying two
completely different methods: the method of hyperspherical harmonics in
momentum representation and the Faddeev equations in configuration space.
Calculations for a binding energy and width of the kaonic three-body system
are performed using three different potentials for the $NN$ interaction, as
well as two different potentials for the description of the kaon-nucleon
interaction. One is the energy independent phenomenological $\bar{K}
N $ potential from Ref. \cite{AYPRC2002}, and the other is the energy
dependent chiral $\bar{K}N$ interaction. Such approach allows one to
understand the dependence of the bound state and width of the kaonic
three-body system on the method of calculations, the importance of
nucleon-nucleon interaction, and the key role of the kaon-nucleon
interactions.

This paper is organized as follows. In Sec. II we describe the effective
potentials of the ${K}^-p$ interaction and present the theoretical
framework to solve the three-body problem for the ${K}^-pp$
system. We apply the method of hyperspherical functions in the momentum
representation for solving the three-body Schr\"{o}dinger equation and the
method of Faddeev equations in configuration space to find bound states for
the kaonic ${K}^-pp$ system$.$ The results of the three-body
calculation for the binding energy and the width for the ${K}^-pp$
cluster and the analysis of the structure of this system are presented in
Sec. 3. We also compare our results with those obtained within variational
methods, the method of hyperspherical functions in the coordinate
representation and the Faddeev equations in momentum representation.
Finally, in the Sec. 4 we present the summary of this work and draw
conclusions.
\section{Theoretical framework}

\subsection{Interactions}

The Hamiltonian of the three nonrelativistic particles for the $\bar{%
K}NN$ system reads %\begin{subequations}
\begin{equation}
H=\widehat{T}+V_{NN}+V_{\bar{K}N_{1}}+V_{\bar{K}N_{2}},
\label{Hamilton}
\end{equation}%
where $\widehat{T}$ \ is the operator of the kinetic energy, $V_{NN}$ is the
nucleon-nucleon potential and $V_{\bar{K}N_{1}}+V_{\bar{K}%
N_{2}}$ is the sum of a pairwise effective antikaon interaction with the
first and second nucleon, respectively. The effective interactions of the $%
\bar{K}N,$ $KN,$ $\bar{K}\bar{K}$ and $K\bar{%
K}$ two-body subsystems are discussed in detail in Refs. \cite{AYPRC2002, AYPL2002, AYKN, HW, DHW, Weise, JidoKanada, KanadaJido}. Below, we use two effective $\bar{K}%
N$ interactions that were derived in different ways. The effective $\overset{%
\_}{K}N$ interactions can be constructed based on a phenomenological
approach so as to reproduce the existing experimental data for the $\overset{%
\_}{K}N$ scattering length, the mass and width of the $\Lambda $(1405)
hyperon and the $1s$ level shift caused by the strong $\bar{K}N$
interaction in the kaonic hydrogen atom. The effective $\bar{K}N$
interaction can be derived within the chiral SU(3) effective field theory,
that presents the low-energy realization of QCD with strange quarks and
identifies the Tomozawa-Weinberg terms as the main contribution to the
low-energy $\bar{K}N$ interaction \cite{Weise}. The 
%single-channel
potential for the description of the $\bar{K}N$ interactions was
derived in Refs. \cite{AYPRC2002, AYKN} phenomenologically using $%
\bar{K}N$ scattering and kaonic hydrogen data and reproducing the $%
\Lambda (1405)$ resonance as a ${K^-}p$ bound state at 1405 MeV.
We refer to this as the Akaishi-Yamazaki (AY) potential. The AY potential is
energy independent. The other $\bar{K}N$ interaction given in Ref. 
\cite{HW} was derived based on the chiral unitary approach for the $s-$wave
scattering amplitude with strangeness $S=-1$ and reproduces the total cross
sections for the elastic and inelastic ${K^-}p$ scattering,
threshold branching ratios, and the $\pi \Sigma $ mass spectrum associated
with the $\Lambda $(1405). We refer to this energy dependent potential for
the parametrization \cite{hwHWJH} as the HW potential and in calculations with this potential, we are following 
a procedure \cite{DHW, KanadaJido}. Both potentials are
local and constructed in coordinate space.

The potentials for both above mentioned effective $\bar{K}N$
interactions can be written in the one-range Gaussian form as

%\end{subequations}
\begin{equation}
V_{\bar{K}N}(r)=\sum\limits_{I=0,1}U^{I}exp\left[ -\left( r/b\right)
^{2}\right] P_{\bar{K}N}^{I},
\end{equation}%
where $r$ is the distance between the antikaon and the nucleon, $b$ is the range
parameter and $P_{\bar{K}N}^{I}$ is the isospin projection operator$%
. $ The values of the potential depth $U^{I=0}$ and $U^{I=1}$ for each \
interaction are given in Refs. \cite{AYPRC2002, AYKN, DHW} and the range parameter is chosen to be $b=0.66$ fm for the AY
potential and $b=0.47$ fm for the HW potential.

%\begin{table}[tbp]
%\caption{Parameters of AY \protect\cite{AYPRC2002} and HW \protect\cite{HW}
%potentials.}
%\label{tab1}%
%\begin{tabular}{ccc|cc}
%\hline\hline
%Interaction & \multicolumn{2}{c|}{AY} & \multicolumn{2}{|c}{HW} \\
%& $U^{I=0},$ MeV & $U^{I=1},$ MeV & $U^{I=0},$ MeV & $U^{I=1},$ MeV \\ \hline
%$\overset{-}{K}N$ & $-595-i83$ & $-175-i105$ & $-908-i181$ & $-415-i170$ \\
%\hline\hline
%\end{tabular}%
%\end{table}
To describe a nucleon-nucleon interaction, we use three $NN$ potentials: the
realistic Argonne V14 (AV14) \cite{ArgonneV14}, the semi-realistic Malfliet
and Tjon MT-I-III (MT) \cite{MT} potential and the Tamagaki G3RS potential 
\cite{Tpotential} that we refer to as the T potential. The T potential was
used in Ref. \cite{AYKN} for variational calculations and that allows us to
compare our results with the latter work. 
%Due to the restriction of our
%consideration by the $s$-wave model for the $\bar{K}N$ interaction,
%the choice for the T potential is reasonable. 
We chose the MT $s$-wave
potential to compare how the binding energy depends on the shape of the $s$-wave $NN$ interaction. 
The MT $s$-wave potential has only two components in
spin-isospin channels $(s,t)$=(1,0), and $(s,t)$= (0,1). The potential
allows one to reproduce and obtain an acceptable description of experimental
data for bound states and scattering in three and four nucleon systems \cite%
{FG1990, CC1998, FY2000}. The realistic AV14 potential perfectly fits both
the $pp$ data, as well as the $np$ data, low-energy $nn$ scattering
parameters and the deuteron properties. This potential was chosen in order
to study the influence of the orbital partial wave components of the $NN$
interaction (by the non central $L^{2}$ operator) with $l>0$ on the energy
of the ${K}^-pp$ cluster. In our considerations, the other
components (spin-orbital, tensor and other) of the realistic potential have
not been taken into account due to their weak effect on the system.
Moreover, in the model with one spin-isospin channel $(s,t)$=(0,1), the
other components of the potential do not make contributions to the
nucleon-nucleon interaction. Thus, the use of all of these potentials allows
the validity test against various $NN$ potentials.

\subsection{Formalism of HH in momentum representation}

Let's introduce the trees of Jacobi coordinates in configuration and
momentum spaces for a system of three particles with unequal masses $%
m_{1},m_{2},$ and $m_{3}$ having positions $\mathbf{r}_{1},\mathbf{r}_{2}$
and $\mathbf{r}_{3}$ and momenta $\mathbf{k}_{1},\mathbf{k}_{2},$ and $%
\mathbf{k}_{3}$\ as follows

\begin{align}
\mathbf{x}_{i}& =\sqrt{\frac{m_{j}m_{k}}{m_{j}+m_{k}}}(\mathbf{r}_{j}-%
\mathbf{r}_{k}),\text{ \ }  \notag \\
\mathbf{y}_{i}& =\sqrt{\frac{m_{j}(m_{j}+m_{k})}{M}}\left( -\mathbf{r}_{i}+%
\frac{m_{j}\mathbf{r}_{j}+m_{k}\mathbf{r}_{k}}{m_{j}+m_{k}}\right) ,  \notag
\\
\mathbf{R}& =(m_{1}\mathbf{r}_{1}+m_{2}\mathbf{r}_{2}+m_{3}\mathbf{r}_{3}),%
\text{ }M=m_{1}+m_{2}=m_{3},\text{ }i\neq j\neq k=1,2,3.  \label{3}
\end{align}%
The conjugate sets of the Jacobi momenta $(\mathbf{q}_{i},\mathbf{p}_{i},%
\mathbf{P})$ for the partition $i$ are defined as:

\begin{align}
\mathbf{q}_{i}& =\frac{1}{\sqrt{m_{j}m_{k}(m_{j}+m_{k})}}(m_{k}\mathbf{k}%
_{j}-m_{j}\mathbf{k}_{k}),\text{ }  \notag \\
\mathbf{p}_{i}& =-\frac{1}{\sqrt{m_{i}M(m_{i}+m_{k}}}[-m_{i}(\mathbf{k}_{j}+%
\mathbf{k}_{k})+(m_{i}+m_{k})\mathbf{k}_{i}],  \notag \\
\mathbf{P}_{i}& =\frac{1}{\sqrt{M}}(\mathbf{k}_{1}+\mathbf{k}_{2}+\mathbf{k}%
_{3}),\text{ }i\neq j\neq k=1,2,3.  \label{JacMomenta}
\end{align}
\bigskip We introduce the hyperspherical coordinates as the hyperradii $%
\varrho $ and $\varkappa ,$ and two sets of five \ angles denoted by$\ \ 
\widetilde{\Omega }_{i}=(\alpha _{i},\widehat{\mathbf{x}}_{i},\widehat{%
\mathbf{y}}_{i})$ and $\Omega _{i}=(\beta _{i},\widehat{\mathbf{q}}_{i},%
\widehat{\mathbf{p}}_{i})$ which define the direction of the vector $\mathbf{%
\rho }$ and vector $\mathbf{\varkappa }$ $,$ in the six dimensional
configuration and momentum spaces, correspondingly, so that 
\begin{align}
\rho & =\sqrt{x_{i}^{2}+y_{i}^{2}},\text{ }x_{i}=\rho \cos \alpha _{i},\text{
}y_{i}=\rho \sin \alpha _{i},\text{ }  \notag \\
d\mathbf{x}_{i}d\mathbf{y}_{i}& =\rho ^{5}d\rho \sin ^{2}\alpha \cos
^{2}\alpha d\alpha d\widehat{\mathbf{x}}_{i}d\widehat{\mathbf{y}}_{i}\equiv
\rho ^{5}d\rho d\widetilde{\Omega }_{i};  \label{Momentum} \\
\varkappa & =\sqrt{p_{i}^{2}+q_{i}^{2}},\text{ }p_{i}=\varkappa \sin \beta
_{i},\text{ }q_{i}=\varkappa \cos \beta _{i},  \notag \\
d\mathbf{p}_{i}d\mathbf{q}_{i}& =\rho ^{5}d\rho \sin ^{2}\beta \cos
^{2}\beta d\beta d\widehat{\mathbf{p}}_{i}d\widehat{\mathbf{q}}_{i}\equiv
\varkappa ^{5}d\varkappa d\Omega _{i}.  \label{Configur}
\end{align}%
The hyperradii $\varrho $ and $\varkappa $ are invariant under
three-dimensional rotations and independent of\text{ }the partition $i$.

One can write the Schr\"{o}dinger integral equation describing three bound
particles in the momentum representation as 
\begin{equation}
\Psi (\mathbf{p},\mathbf{q})=-\frac{1}{(2\pi )^{6}}\int G(\mathbf{p},\mathbf{%
q})<\mathbf{p}^{^{\prime }}\mathbf{q}^{^{\prime }}\left\vert
V_{123}\right\vert \mathbf{pq>}\Psi (\mathbf{p}^{^{\prime }},\mathbf{q}%
^{^{\prime }})d\mathbf{p}^{\prime }d\mathbf{q}^{\prime },
\label{Schrodinger_Momentum}
\end{equation}%
where 
\begin{equation}
<\mathbf{p}^{^{\prime }}\mathbf{q}^{^{\prime }}\left\vert V_{123}\right\vert 
\mathbf{pq>}=\frac{1}{(2\pi )^{6}}\int V_{123}\exp \left[ i(\mathbf{q-q}%
^{^{\prime }})\mathbf{x+}i(\mathbf{p-p}^{^{\prime }})\mathbf{y}\right] d%
\mathbf{x}d\mathbf{y},  \label{Potential_Momentum}
\end{equation}%
is the Fourier transformation of the $V_{123},$ which is defined as the sum
of the pair-wise nucleon-nucleon and effective antikaon-nucleon
interactions. In Eq. (\ref{Schrodinger_Momentum}), the Green function has
the form 
\begin{equation}
G(\mathbf{p},\mathbf{q})=\frac{2M}{\hslash^{2}}\frac{1}{p^{2}+q^{2}+%
\varkappa_{0}^{2}},  \label{Green Func}
\end{equation}
where $\varkappa_{0}^{2}=\frac{2ME}{\hslash^{2}},$ and $E$ is the bounding
energy of the three-particle system.

For the next step, we follow the procedure given in Refs. \cite
{JibutiNuclPhys, DzhbutiKrup} and expand the wave function of three bound
particles in terms of the antisymmetrized hyperspherical harmonics $\Phi _{\mu
}^{l_{p}l_{q}}(\Omega _{\varkappa },\mathbf{\sigma },\mathbf{\tau })$ in the
momentum representation: 
\begin{equation}
\Psi (\varkappa ,\widetilde{\Omega })=\sum\limits_{\mu l_{p}l_{q}}u_{\mu
}^{l_{p}l_{q}L}(\varkappa )\Phi _{\mu }^{l_{p}l_{q}L}(\Omega _{\varkappa },%
\mathbf{\sigma },\mathbf{\tau }),  \label{HH3}
\end{equation}%
where $\mu $ is the grand angular momentum, $L$ is total orbital momentum
and $l_{p}$ and $l_{q}$ are the angular momenta corresponding to the momenta 
$p$ and $q$. In Eq. (\ref{HH3}) the antisymmetrized hyperspherical functions $\Phi _{\mu }^{l_{p}l_{q}L}(\Omega _{\varkappa },\mathbf{%
\sigma },\mathbf{\tau })$ are written as a sum of products of spin and isospin functions and
hyperspherical functions $\Phi _{\mu }^{l_{p}l_{q}L}(\Omega
_{i})=\sum\limits_{m_{p}m_{q}}\left\langle l_{p}l_{q}m_{p}m_{q}\right\vert
\left. LM\right\rangle \Phi _{\mu }^{l_{p}l_{q}m_{p}m_{q}}(\Omega
_{i})$, where $\Phi _{\mu }^{l_{p}l_{q}m_{p}m_{q}}(\Omega
_{i})$ are hyperspherical harmonics, using the Raynal-Revai
coefficients \cite{RaynalRevai}. The HH $\Phi _{\mu
}^{l_{p}l_{q}m_{p}m_{q}}(\Omega _{\varkappa })$ are the eigenfunctions of
the angular part of the six-dimensional Laplace operator in momentum space
with eigenvalue $\mu (\mu +4)$, and are expressible in terms of spherical
harmonics and Jacobi polynomials \cite{DzhbutiKrup, Avery}$.$ By
substituting Eq. (\ref{HH3}) into the integral Schr\"{o}dinger equation in
the momentum representation (\ref{Schrodinger_Momentum}) and taking into
account (\ref{Green Func}) one obtains a system of coupled integral
equations for the hyperradial functions\ $u_{\mu }^{l_{p}l_{q}L}(\varkappa
): $ 
\begin{align}
(\varkappa ^{2}+\varkappa _{0}^{2})u_{_{\mu }}^{l_{q}l_{p}L}(\varkappa )& =-%
\frac{2m}{\hbar ^{2}}\frac{1}{\varkappa ^{2}}\underset{\mu ^{\prime }i}{\sum 
}\sum_{l_{p}^{^{\prime }}l_{q}^{^{\prime }}\bar{l}_{p}\bar{l}%
_{q}A}\text{ }^{j}\langle \bar{l}_{p}\bar{l}%
_{q}|l_{p}l_{q}\rangle _{\mu L}^{i}\text{ }  \notag \\
& ^{j}\langle \bar{l}_{p}\bar{l}_{q}|l_{p}^{^{\prime
}}l_{q}^{^{\prime }}\rangle _{\mu ^{^{\prime }}L}^{i}\int \rho d\rho J_{\mu
+2}(\varkappa \rho )J_{\mu ^{^{\prime }}+2}(\varkappa ^{^{\prime }}\rho
)\times  \notag \\
& \Phi _{\mu }^{\bar{l}_{p}\bar{l}_{q}}\left( \alpha
_{_{i}}\right) V_{A}(x_{i})\Phi _{\mu ^{^{\prime }}}^{\bar{l}_{p}%
\bar{l}_{q}}\left( \alpha _{_{i}}\right) u_{_{\mu ^{\prime
}}}^{l_{q}^{\prime }l_{p}^{\prime }L}(\varkappa ^{\prime })d\varkappa
^{\prime 3}d\Omega _{j},  \label{HH Intergral Eq.}
\end{align}%
where the index $A$ is related to the type of interaction $A\in NN$, $%
\bar{K}N.$ In Eq. (\ref{HH Intergral Eq.}) $^{j}\langle \bar{%
l}_{p}\bar{l}_{q}|l_{p}l_{q}\rangle _{\mu L}^{i}$ are the
Raynal-Revai coefficients \cite{RaynalRevai} for the unitary transformation
of HH from one set of Jacobi coordinate to another, $J_{\mu +2}(\varkappa
\rho )$ are the spherical Bessel functions and 
\begin{align}
\Phi _{\mu }^{l_{p}l_{q}}(\alpha _{i})& =N_{\mu }^{l_{p}l_{q}}(\cos \alpha
_{i})^{l_{p}}(\sin \alpha _{i})^{l_{q}}P_{n}^{l_{p}+1/2,l_{q}+1/2}(\cos
2\alpha _{i}),  \notag \\
N_{\mu }^{l_{p}l_{q}}\ & =\sqrt{\frac{2n!(\mu +2)(n+l_{p}+l_{q}+1)!}{\Gamma
(n+l_{1}+3/2)\Gamma (n+l_{2}+3/2)}},
\end{align}%
where $2n=\mu -l_{q}-l_{p}$, $n$=1, 2, 3,... and $P_{n}^{l_{p}l_{q}}$ is the Jacobi
polynomial. By solving the coupled integral equations (\ref{HH Intergral Eq.}%
) one can find the hyperradial functions $u_{_{\mu
}}^{l_{q}l_{p}L}(\varkappa )$ for a given $L.$

\subsection{Formalism of Faddeev equation in configuration space}

The wave function of the three--body system can be obtained by solving the
Schr\"{o}dinger equation with the single channel  Hamiltonian (\ref{Hamilton}).
Alternatively, in the Faddeev method the total wave function can be
decomposed into three components: $\Psi =\Phi _{1}+\Phi _{2}+\Phi _{3}$ \cite%
{FaddeevConfigurSpace}. The Faddeev components $\Phi _{i}$ correspond to
the separation of particles into configurations $i+(kl)$, $i\neq k\neq
l=1,2,3$. Each Faddeev component $\Phi _{i}=\Phi _{i}(\mathbf{x}_{i},\mathbf{%
y}_{i})$ depends on its own set of the Jacobi coordinates (\ref{3}). The
components satisfy the Faddeev equations in the coordinate representation
written in the form \cite{FaddeevConfigurSpace, NF68, K86}: 
\begin{equation}
\left( H_{0}^{i}+v_{i}(\mathbf{x_{i}})-E\right) \Phi _{i}(\mathbf{%
x_{i},y_{i})}=-v_{i}(\mathbf{x_{i}})\left( \Phi _{k}(\mathbf{x_{k},y_{k}}%
)+\Phi _{l}(\mathbf{x_{l},y_{l}})\right) ,
\label{fadnoy}
\end{equation}%
where $ i\neq k\neq l=1,2,3$.  $H_{0}^{i}=-\frac{\hbar ^{2}}{M_{i}}(\Delta _{\mathbf{x}_{i}}+\Delta _{%
\mathbf{y}_{i}})$ is the kinetic energy operator, $M_{i}$ is the reduced
mass for the corresponding Jacobi coordinates (\ref{3}) and $v_{i}$ is the
potential acting between the particles $(kl)$, $i\neq k\neq l$. The
orthogonal transformation between three different sets of the Jacobi
coordinates has the form: 
\begin{equation}
\left( 
\begin{array}{c}
\mathbf{x}_{i} \\ 
\mathbf{y}_{i}%
\end{array}%
\right) =\left( 
\begin{array}{cc}
C_{ik} & S_{ik} \\ 
-S_{ik} & C_{ik}%
\end{array}%
\right) \left( 
\begin{array}{c}
\mathbf{x}_{k} \\ 
\mathbf{y}_{k}%
\end{array}%
\right) ,\ \ C_{ik}^{2}+S_{ik}^{2}=1,  \label{difjacobi}
\end{equation}
where 
$$
C_{ik}=-\sqrt{\frac{m_{i}m_{k}}{(M-m_{i})(M-m_{k})}},\text{ }
S_{ik}=(-1)^{k-i}\mathrm{sign}(k-i)\sqrt{1-C_{ik}^{2}}.
$$

The bound state of the ${K}^-pp$ system can be obtained by solving
the Faddeev equations in configuration space (\ref{fadnoy}). In this case
the total wave function of the ${K}^-pp$ system is decomposed into
the sum of the Faddeev components $U$ and $W$ corresponding to the $(pp)
{K}^-$ and $({K}^-p)p$ types of rearrangements: $\Psi
=U+W-PW$, where $P$ is the permutation operator for two identical
fermions. For a three--body system that includes two identical fermions the
set of the Faddeev equations (\ref{fadnoy}) can be reduced to the system of
two equations for the components $U$ and $W$ \cite{14,14a}: 
\begin{equation}
\begin{array}{l}
{(H_{0}^{u}+V_{pp}-E)U=-V_{pp}(W-PW),} \\ 
{(H_{0}^{w}+V_{{K}^-p}-E)W=-V_{{K}^-p}(U-PW),}%
\end{array}
\label{GrindEQ__1_}
\end{equation}%
where the potentials for $pp$ and ${K}^-p$ pairs are defined by $%
V_{pp}$ and $V_{{K}^-p}$, respectively. The partial wave analysis
of the differential Faddeev equations (DFE) (\ref{GrindEQ__1_}) in the $LS$
basis is performed by the general scheme described in Refs. \cite{K86,14,15}. 
The $LS$ basis allows us to restrict the model space to the states with
total spin $S=0$ (when the spin projections of protons are anti-parallel)
and total isospin $I=\frac{1}{2}$. The possible isospin configurations with $%
I=1$ are not taken into account in our calculations. According to the
evaluations of different authors the total contribution of the $I$=1
configuration is about 5\%~\cite{AYKN}. We consider two cases for the
solution of the system (\ref{GrindEQ__1_}): the $s$-wave approach and the
cluster approach. In the latter case we assume the dominant clustering of
the ${K}^-pp$ system in the form $\Lambda (1405)+p$, that allows
us to present the Faddeev components as the product of the eigenfunctions of
the two-particle subsystems and functions of the relative motion of pair and
the third particle.

\subsubsection{$s$-wave approach}

\begin{figure}[h]
\centering
\includegraphics[scale=.225]{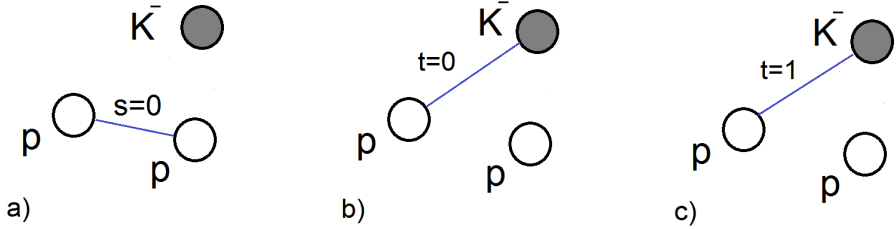}
\caption{ Spin-isospin configurations of the subsystems of the $
\protect{K^-}pp$ system. a) ${K}^-$+$(pp)$ with
the singlet spin state of the pair $(pp)$. b) $(\protect{K^-}p)+p$
with the singlet isospin state of the pair $(\protect{K^-}p)$. c) $
(\protect{K^-}p)+p$ with the triplet isospin state of the pair $(
\protect{K^-}p)$. }
\label{fig1}
\end{figure}

Let us consider the $s$-wave approach. Above, we mentioned that the
nucleon-nucleon MT and T potentials and ${K}^-p$ interactions
applied have only the $s$-wave components for the spin-isospin state $s$=0, $%
t$=1 in the pair $NN$ and $t$=0,1 in the pair ${K}^-p$. We
consider the $s$-wave approach for the differential Faddeev equations (\ref%
{GrindEQ__1_}). The $s$-wave DFE can be written as 
$$
(H_{0}^{u}+v_{pp}^{s}(x)-E)U(x,y)=-v_{pp}^{s}(x)\int\limits_{-1}^{1}du\frac{%
xy}{x_{1}^{\prime }y_{1}^{\prime }}AW(x_{1}^{\prime },y_{1}^{\prime }),
$$
\begin{equation}
(H_{0}^{w}+V_{{K}^-p}(x)-E)W(x,y)=-\frac{1}{2}V_{{K}^-
p}(x)\left[ \int\limits_{-1}^{1}du\frac{xy}{x_{2}^{\prime }y_{2}^{\prime }}%
A^{T}U(x_{2}^{\prime },y_{2}^{\prime })+\right.  \label{DFE1}
\end{equation}%
$$
\left. +\int\limits_{-1}^{1}du\frac{xy}{x_{2}^{\prime \prime }y_{2}^{\prime
\prime }}DW(x_{2}^{\prime \prime },y_{2}^{\prime \prime })\right] ,
$$
where the numbers of particles in the system are labeled as 2 and 3 for
protons and 1 for the kaon, $m_{2}=m_{3}=m$ is the nucleon mass, and $m_{1}$
is the mass of the kaon. $v_{pp}^{s}(x)$ is the singlet-triplet ($s$=0 and $%
t $=1) component of the $NN$ potential, $V_{{K}^-p}=$diag$\{v_{%
{K}^-p}^{s},v_{{K}^-p}^{t}\}$ with $v_{{K}^-p
}^{s}$, and $v_{{K}^-p}^{t}$ the singlet and triplet isospin
components of $\bar{K}N$ potential, respectively. In Eqs. (\ref{DFE1}%
) $u$=$cos(\widehat{\mathbf{xy}})$, where $\widehat{{\mathbf{xy}}}$ is the angle between $\mathbf{x}$ and $\mathbf{y}$,
and we define 
\begin{align}
H_{0}^{u}& =-\frac{\hbar ^{2}}{2M^{(1)}}\partial _{y}^{2}-\frac{\hbar ^{2}}{%
2m^{(1)}}\partial _{x}^{2},\qquad H_{0}^{w}=-\frac{\hbar ^{2}}{2M^{(2)}}%
\partial _{y}^{2}-\frac{\hbar ^{2}}{2m^{(2)}}\partial _{x}^{2},  \notag \\
A& =(-\frac{\sqrt{3}}{2},-\frac{1}{2}),\qquad D=\left( 
\begin{array}{cc}
\frac{1}{2} & \frac{\sqrt{3}}{2} \\ 
\frac{\sqrt{3}}{2} & -\frac{1}{2}%
\end{array}%
\right) ,  \label{DFE2} \\
W(x,y)& =(W^{s}(x,y),W^{t}(x,y))^{T}.  \notag
\end{align}
In (\ref{DFE1}), the appropriate transformation of coordinates and reduced masses are given by 
\begin{equation}
x_{1}^{\prime }=\left( (\frac{m^{(1)}}{m}x)^{2}+y^{2}+2\frac{m^{(1)}}{m}%
xyu\right) ^{1/2},\text{ \ \ \ }y_{1}^{\prime }=\frac{m_{1}}{m_{1}+m}\left( (%
\frac{m}{M^{(1)}}x)^{2}+y^{2}-2\frac{m}{M^{(1)}}xyu\right) ^{1/2},
\end{equation}%
$$
x_{2}^{\prime }=\left( (\frac{m^{(2)}}{m}x)^{2}+y^{2}+2\frac{m^{(2)}}{m}%
xyu\right) ^{1/2},\text{ \ \ }y_{2}^{\prime }=\frac{1}{2}\left( (\frac{m}{%
M^{(2)}}x)^{2}+y^{2}-2\frac{m}{M^{(2)}}xyu\right) ^{1/2},
$$
$$
x_{2}^{\prime \prime }=\left( (\frac{m^{(2)}}{m_{1}}x)^{2}+y^{2}-2\frac{%
m^{(2)}}{m_{1}}xyu\right) ^{1/2},\text{ \ }y_{2}^{\prime \prime }=\frac{m}{%
m_{1}+m}\left( (\frac{m_1}{M^{(2)}}x)^{2}+y^{2}+2\frac{m_{1}}{M^{(2)}}%
xyu\right) ^{1/2}.
$$
$$
M^{(1)}=2\frac{m_{1}m}{m_{1}+2m},\qquad M^{(2)}=\frac{m(m_{1}+m)}{m_{1}+2m},
$$
where $ m^{(1)}=\frac{m}{2}$,$\quad$ $m^{(2)}=\frac{m_{1}m}{m_{1}+m}$ are reduced masses
the $pp$ pair and the pair and third particle ($K^-$).
Spin-isospin configurations of the subsystems of the ${K^-}pp$
system, corresponding to (\ref{DFE1}) are graphically presented in Fig. \ref{fig1}. 

\subsubsection{Averaged potential}

The $\bar{K}$ meson combines two nucleons into the bound state for
two isospin configurations that are energetically favorable. The effective $
\bar{K}N$ interactions have strong attraction for the $I=0$ channel
and weak attraction in the $I=1$ channel. In this section, we consider an
approach for the case when one of the pair potentials has different
components for spin-isospin states. Following, for instance Refs. \cite{15}
and \cite{KanadaJido}, for Eqs. (\ref{HH Intergral Eq.}) and (\ref{DFE1}),
one can consider an effective potential obtained by averaging the initial
potential over isospin variables. Each component of the initial potential is
substituted by the averaged potential. The isospin averaged potential $V_{
{K}^-p}(r)$ is defined by the following form: 
\begin{equation}
V_{{K}^-p}(r)=\frac{3}{4}v_{{K}^-p}^{I=0}(r)+\frac{1}{4}
v_{{K}^-p}^{I=1}(r).  \label{av_pot}
\end{equation}%
This potential has moderate attraction in comparison with the component for $%
{I=0}$. Using this potential we reduce, for example, the set of equations (%
\ref{DFE1}) to two equations. The averaged potential replaces the $V_{
{K}^-p}$ so that $$V_{{K}^-p}=diag\{v_{{K}^-
p}^{av.},v_{{K}^-p}^{av.}\}$$ and $\mathcal{W}(x,y)$ becomes $%
\mathcal{W}(x,y)=-\frac{\sqrt{3}}{2}W_{1}-\frac{1}{2}W_{2}$. Note that this
simplification changes the two-body threshold in Eq. (\ref{GrindEQ__1_}),
which is not related to the $V_{{K}^-p}$ bound state as $\Lambda $%
(1405). The value of the two-body threshold for the effective potential is
about -9.6~MeV instead of -30.2~MeV for the AY potential ($I=0)$.

\subsubsection{Cluster approach}

%%%%%%%%%%%%%%%%%%%%%%%%%%%%%%%%%%%%%%%%%%%%%%%%%%%%%%%%%%%%%%%
Within the formalism of the Faddeev equations, one may use dominant
clustering of the ${K^-}pp$ system in the form $\Lambda (1405)+p$
to calculate the binding energy. In the framework of such an approach, the $%
s $-wave Faddeev components are decomposed as a product of the
eigenfunctions of the Hamiltonians of two-particle subsystems and functions
of the relative motion of the pair and the third particle.

Here, we present a brief description of the method based on the general form
for the Faddeev equations (\ref{fadnoy}). The Faddeev components $\mathcal{U}
_{\alpha }$, $\alpha $=1,2,3 are written in the following form \cite{F2000}: 
\begin{equation}
\mathcal{U}_{\alpha }(x,y)=\phi _{\alpha }(x)f_{\alpha }(y).  \label{exp}
\end{equation}

In (\ref{exp}), the functions $\phi _{\alpha }$ are the solutions of the
two-body Schr\"{o}dinger equations for subsystems with minimal eigenvalue 
$$
\left( -\frac{\hbar ^{2}}{2m^{(\alpha )}}\partial _{x}^{2}+v_{\alpha
}(x)\right) \phi _{\alpha }(x)=\varepsilon _{\alpha }\phi _{\alpha }(x).
$$
Substituting (\ref{exp}) into the Faddeev equations and projecting, one may
obtain the set of integro-differential equations for the functions $
f_{\alpha }(y)$ describing the relative motion of clusters: 
\begin{align}
& \left( -\frac{\hbar ^{2}}{2M^{(\alpha )}}\partial _{y}^{2}+\varepsilon
_{\alpha }-E\right) f_{\alpha }(y)  \notag \\
& =-\frac{1}{2}\left\langle \phi _{\alpha }(x)\Big |v_{\alpha
}(x)\int\limits_{-1}^{1}du\left\{ \frac{xy}{x_{\beta }^{\prime }y_{\beta
}^{\prime }}\phi _{\beta }(x_{\beta }^{\prime })f_{\beta }(y_{\beta
}^{\prime })\right. \right. +\left. \left. \frac{xy}{x_{\gamma }^{\prime
\prime }y_{\gamma }^{\prime \prime }}\phi _{\gamma }(x_{\gamma }^{\prime
\prime })f_{\gamma ,k}(y_{\gamma }^{\prime \prime })\right\} \right\rangle .
\label{E7}
\end{align}%
In these equations, $\langle .|.\rangle $ means the integration over the
variables $x$, and the indexes $\alpha \neq \beta \neq \gamma =1,2,3$. The
functions $f_{\alpha }(y)$ satisfy the following boundary conditions $
f_{\alpha }(y)\sim 0,$ when $y$\ $\rightarrow $\ $\infty $\emph{.} For a
system including particles with spin and isospin, the number of equations in
the set depends on the number of terms of the Faddeev components (\ref{exp})
in the spin-isospin basis. For the ${K^-}pp$ system considered within the framework of the $s$%
-wave approach, we have the set of three equations corresponding to (\ref
{DFE1}). The set of eigenvalues of pair subsystems $\epsilon _{i}$, $i=1,2,3$
includes $\epsilon _{2}$ as the bound state energy of the ${K^-}p$
pair in singlet isospin state assumed to be the $\Lambda (1405)$ hyperon. 
%\begin{figure}[h]
%\centering
%\includegraphics[scale=.45]{Fig3.eps}
%\caption{ The function of relative motion within the cluster approach for $%
%K^{-}pp$ system: a) $K^{-}+(pp)$ b) $(K^{-}p)+p$ with singlet isospin state
%of the pair $(K^{-}p)$ c) $(K^{-}p)+p$ with triplet isospin state of the
%pair $(K^{-}p)$. The configurations are related to correspond equation in
%the set Eq.(2). }
%\label{fig3}
%\end{figure}

\section{Results and discussion}

\begin{figure}[tbp]
\begin{center}
\includegraphics[width = 8.75cm]{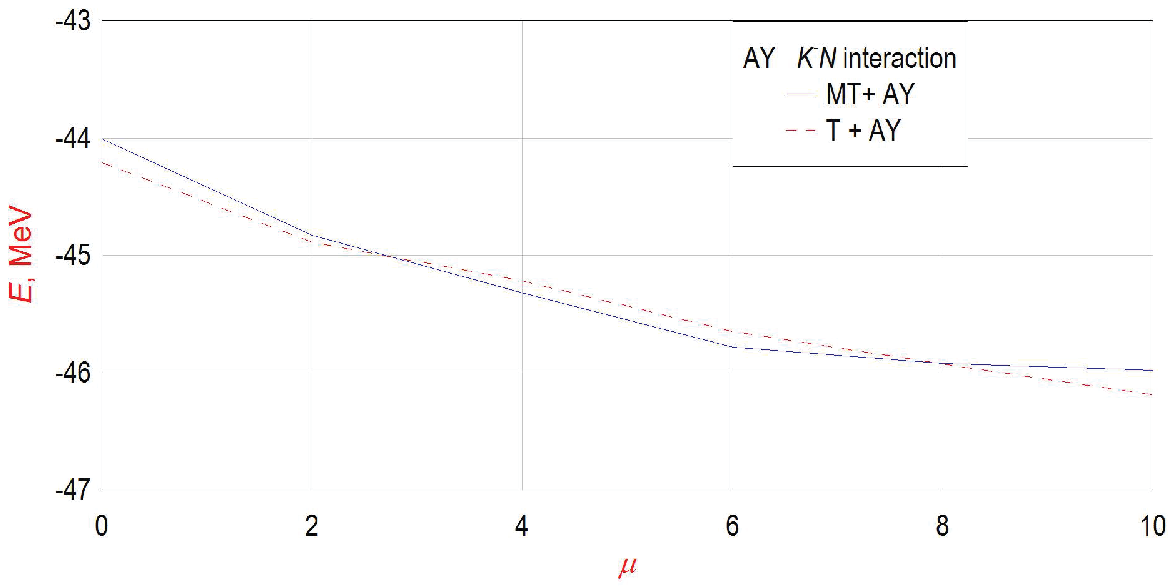}%\hspace{0.25cm}
\includegraphics[width = 8.75cm]{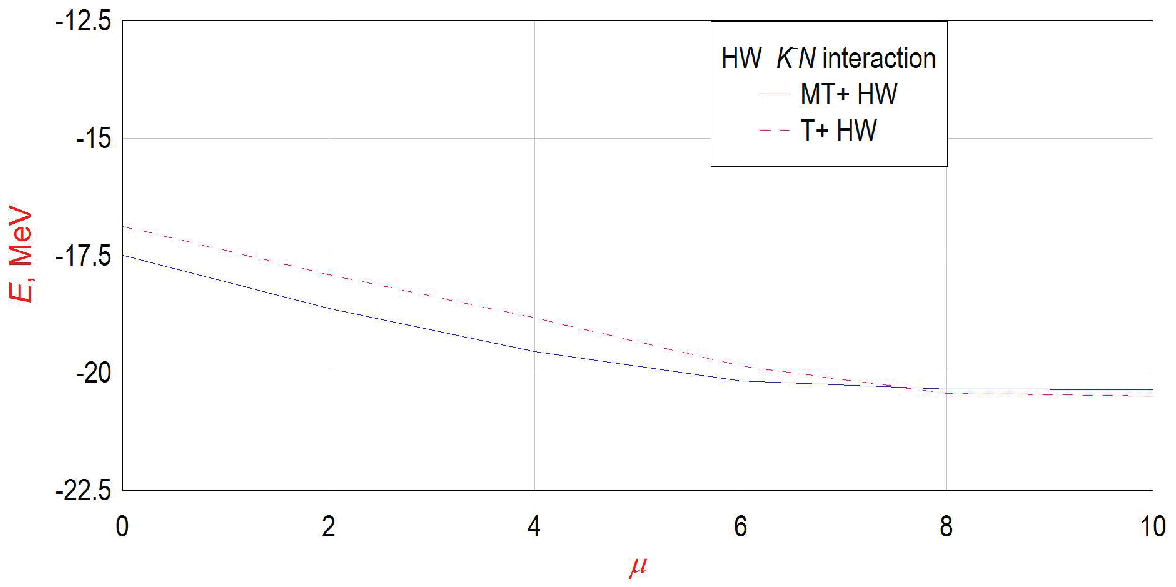}
\end{center}
\caption{The convergence of the ground state energies $E$ of the $%
\protect{K^-}pp$ system as a function of $\mu _{max}$   for different $NN$ potentials and $%
\protect{K^-}N$ interactions.}
\label{fig2}
\end{figure}

Results of our calculations for the ${K^-}pp$ cluster are
presented in this section. For the calculations of the binding energy and
the width with the methods of HH and DFE we use for the $NN$ interaction
MT, T and AV14 potentials, while for the $\bar{K}N$ interaction we
use the single-channel phenomenological AY $\bar{K}N$ potential with the
range parameter $b$ = 0.66 fm and the energy dependent effective HW potential with the range parameter $b$=0.47 fm. Such an approach allowed us to examine how
the ${K^-}pp$ cluster's structure depends on different choices of
the $\bar{K}N$ interactions for the same $NN$ potential, as well as
dependence from different choices of the $NN$ interaction for the same $%
\bar{K}N$ interaction. This enable us to understand the sensitivity
of the system to the input interactions. By solving the system of equations (%
\ref{HH Intergral Eq.}) one finds the 
binding energy of the ${K^-}pp$ cluster. The results
of the calculations with the method of HH strongly depend on the number of
terms in the expansion of the wave function (\ref{HH3}). The convergences of
binding energy calculations for the ground state of the ${K^-}pp$
system as a function of the grand angular momentum $\mu$  are shown in Fig. \ref{fig2} for
different input potentials. Reasonable convergence is reached for $\mu
_{max}=10$ and we limit our considerations to this value. As is seen from
Fig. \ref{fig2} much faster convergence occurs for considered $NN$
potentials when the AY potential is chosen for the $\bar{K}N$
interaction, while for the HW $\bar{K}N$ interaction the binding
energy calculations converge more slowly. The accuracy reached is about 0.2
MeV and our expectation is that consideration of the higher values of the
grand angular momentum $\mu $ would not dramatically change the binding
energy of the ${K^-}pp$ cluster. 

\begin{table}[t]
\begin{center}
\caption{The binding energy $B$ and width $\Gamma $ for the $\protect{K^-}pp$ system calculated in the framework of the method of HH in the
momentum representation for different interactions. $NN$ potentials: MT \protect\cite{MT} and T \protect\cite%
{Tpotential}. $\protect\bar{K}N$ interactions: AY \protect\cite%
{AYPRC2002} and HW \protect\cite{HW}. $E_{\protect{K^-}p}$ is
two-body energy of the $\protect{K^-}p$.}
\label{tab2}%
\begin{tabular}{c|cccc}
\hline\hline
& MT+AY & T+AY &\multicolumn{1}{|c} MT+HW & T+HW \\ \hline
$B$, MeV & 46.5 & 46.3 & \multicolumn{1}{|c}{20.5} & 20.6 \\ 
$\Gamma $, MeV & 84.3 & 74.5 & \multicolumn{1}{|c}{48.1} & 49.5 \\ \hline
$E_{{K^-}p},$ MeV & \multicolumn{2}{|c}{29.9} & 
\multicolumn{2}{|c}{10.9} \\ \hline\hline
\end{tabular}
\end{center}
\end{table}
The results for the binding energy and the width of the ${K^-}pp$
system using the method of HH for different $\bar{K}N$ \ and $NN$
interactions are presented in Table \ref{tab2}. $E_{{K^-}p}$ is
the two-body energy when only the ${K^-}p$ pair are interacting,
while the interaction between the two protons is neglected. The HH method
allows one to obtain the wave function of the ${K^-}pp$ cluster.
By solving the system of equations (\ref{HH Intergral Eq.}) one finds the
binding energy as well as the corresponding hyperradial functions. The
latter allows one to construct the wave function $\Psi $ for the ${K}^-pp$ system. Using the wave function, the width of the bound state can be
evaluated in a perturbative way from the imaginary part of the $\bar{K}N$ interaction as 
$\Gamma =-$2 $\left\langle \Psi \left\vert \text{Im}
\left( V_{{K^-}p}(r_{12})+V_{{K^-}p}(r_{13})\right)
\right\vert \Psi \right\rangle $. Since $\left\vert \text{Im}V_{{K^-}p}(r)\right\vert \ll \left\vert \text{Re}
V_{{K^-}p}(r)\right\vert $, this is a reasonable approximation for the width. As it is
stated in Gal's review \cite{Gal}, as well as demonstrated in the recent
calculations of the width for the ${K^-}pp$ system \cite{Dote2015} using a coupled-channel complex
scaling method with Feshbach projection, this is a good approximation. For an approximate evaluation of the width the imaginary part of the complex
potential has often been treated perturbatively by many authors \cite
{ADYPL2005, DotePRC2004, DW2007, DHWnp2008, DHW, WycechGreen, BarneaK, RKSh.Ts2014, KanadaJido}. Recently this approach was used in Ref. \cite{Barnea2015} 
for few-body calculations of $\eta$-nuclear quasibound states. We use the latter
expression to find the width and the calculated values of the width for
different $\bar{K}N$ \ and $NN$ interactions are listed in Table \ref{tab2}. The analysis of the
results presented in Table \ref{tab2} shows that, for the chiral HW $\bar{K}N$ interaction, the binding energy is less than half that of the
phenomenological AY interaction and the HW interaction also leads to a much
smaller width. However, the binding energy and the width are not sensitive
to the form of $NN$ potentials for the same type of $\bar{K}N$
interaction. The binding energy turns out to be about 1 MeV, while the width
ranges from $\sim $75 to $\sim $84 MeV for the AY potential and turns out to
be within about 1.5 MeV for the HW $\bar{K}N$ interaction. 
\begin{table}[t]
\begin{center}
\caption{\label{tab3} Orbital momentum configurations of $(pp)\protect{K^-}$
and $(\protect{K^-}p)p$ rearrangements and the ground state energy
of the $\protect{K^-}pp$ system calculated with the HW and AY $%
\protect\bar{K}N$ potentials. The $NN$ interaction is given by the
AV14 potential \protect\cite{ArgonneV14}. $l$ is the orbital momentum of the
pair of particles, and $\protect\lambda $ is the orbital momentum associated
with the relative motion of the third particle with  respect of the center of mass
of the pair.}%
%\begin{tabular}{p{1.in}p{0.3in}p {1.0in}p{1.1in}p{1.in}p{1.in}} \hline%
\begin{tabular}{clccc}
\hline\hline
& {$(pp)\protect{K^-} $} & {$(\protect{K^-}p)p$} & HW & AY \\ \hline
{($l$, $\lambda$)} & {(0,0) } & {(0,0) } & {-21.15} & {-46.97} \\ 
{($l$, $\lambda$)} & {(0,0) (2,2) } & {(0,0) } & {-21.54} & {-47.33} \\ 
{($l$, $\lambda$)} & {(0,0) (2,2) (4,4)} & {(0,0) } & {-21.56} & {-47.34} \\ \hline\hline
\end{tabular}
\end{center}
\end{table}

We will now discuss the results when we employ the Faddeev technique. The
results for the binding energy of the ${K^-}pp$ system for the
orbital momentum configurations of $(pp){K^-}$ and $({K^-}p)p$ rearrangements are presented in Table \ref{tab3}. The calculations are
performed using the AV14 $NN$ potential for the AY and HW $\bar{K}N$
interactions. As can be seen from Table \ref{tab3}, the orbital
contributions with ($l,\lambda )$ are equal $(2,2)$ and  $(4,4)$ are small
enough and the $s-$wave consideration is very reasonable. The numerical
convergence for the binding energy calculations is fast with increasing
model space. The visible contribution that provides the bound state of the
system comes from the $p$-wave of the $pp$ pair. For the AV14 potential,
we also calculated the binding energy using the averaged potential approach
as well as the cluster approach. The results with the averaged potential (%
\ref{av_pot}) are 33.7 MeV and 12.5 MeV for the AY and HW $\bar{K}N$
interactions, respectively. 
%Therefore, these results are in good agreement
%with those obtained with the averaged potential (\ref{av_pot}) using the
%method of HH for the same $NN$ interaction. 
\begin{figure}[t]
\centering
\includegraphics[scale=.5]{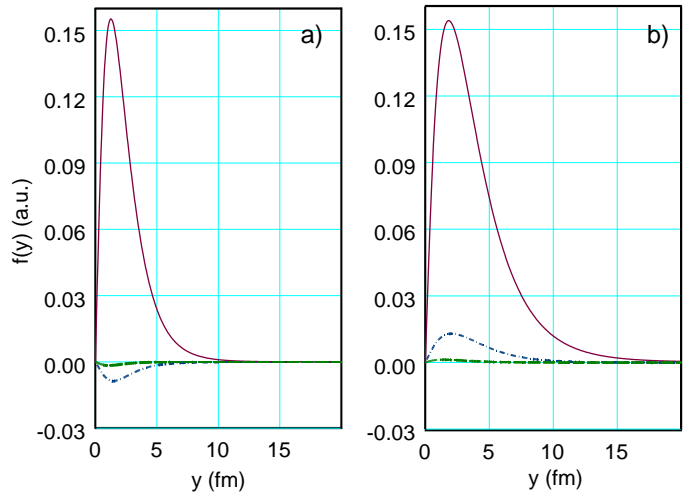}
\caption{ The function of relative motion within the cluster approach for
the $\protect{K^-}pp$ system: the dot-dashed curve shows
calculation for the configuration $\protect{K^-}+(pp);$ the solid
curve shows calculation for the configuration $(\protect{K^-}p)+p$
with the singlet isospin state of the pair $(\protect{K^-}p);$ the
dashed curve shows calculation for the configuration $(\protect{K^-}
p)+p$ with the triplet isospin state of the pair $(\protect{K^-}p)
$. The configurations are related to the set of Eqs. (\protect\ref{E7}). The
calculations are performed with the nucleon-nucleon MT interaction for a)
the AY effective $\protect\bar{K}N$ potential and b) the HW
effective $\protect\bar{K}N$ potential. }
\label{fig4}
\end{figure}
We used dominant clustering of the ${K^-}pp$ system in the
configuration $\Lambda (1405)+p$ to calculate the binding energy within the
cluster approach for the DFE. 
The Faddeev components are decomposed as a product of the eigenfunctions of
the Hamiltonians for the two particle subsystems and the function of the
relative motion of the pairs and the third particle. Our results for the
functions associated with the relative motion are shown in Fig.~\ref{fig4}
for the nucleon-nucleon MT potential and the AY and HW effective $\bar
{K}N$ interactions.  
One can evaluate the dominant contribution of the $%
({K^-}p)^{singlet}+p$ configuration to the total wave function of the
system, which is represented as a sum of the Faddeev components. The calculations for 
other $NN$ potentials have shown the same behavior and relative contributions 
of the functions associated with relative motion. In the
cluster approach, the binding energy increases for both ${K^-}p$
interactions and are 49.5 MeV and 17.7 MeV, respectively. 

\begin{table}[tbp]
\begin{center} 
\caption{The proton separation energy $E_{(\protect{K^-}p)-p}$
calculated for the MT $NN$ potential and the HW and AY $\protect\bar{%
K}N$ interactions. Energies are given in unit of MeV.}
\label{tab4a}%
\begin{tabular}{ccccc}
\hline\hline 
$\bar{K}N$ & Averaged potential & Cluster & DFE &  \\ 
interaction & approach & approach &  &  \\ \hline
AY & 24.1 & 19.2 & 15.7 &  \\ 
HW & 11.2 & 6.5 & 9.3 &  \\ \hline\hline
\end{tabular}
\end{center} 
\end{table}
Comparing the binding energies obtained by using the averaged potential and
the cluster approaches with the $s$-wave DFE calculations, one can conclude
that the averaged potential approach gives the better result for the
relatively weaker HW potential. At the same time, the cluster approach is
more accurate for the stronger AY potential. We also calculated the proton
separation energy within the averaged potential, the cluster and the $s$-wave DFE approaches. 
The results are presented in Table \ref{tab4a}.

The summary of the results for the binding energy of the ${K}^-pp$
system in the framework of the method of DFE for all considered $NN$ and $%
\bar{K}N$ interactions are presented in Table \ref{tab4}. 
\begin{table}[t]
\begin{center} 
\caption{The binding energy $B$ for the ${K}^-pp$ system
calculated in the framework of the DFE method for different $NN$ and $%
\protect\bar{K}N$ interactions. $NN$ potentials: AV14 \protect\cite%
{ArgonneV14}, MT \protect\cite{MT} and T \protect\cite{Tpotential}. $%
\protect\bar{K}N$ interactions: AY \protect\cite{AYPRC2002} and HW 
\protect\cite{HW}. \textit{a} is the two-body scattering length and $%
\left\langle r^{2}\right\rangle ^{1/2}$ is the root-mean-squared distance in
the $\protect{K^-}p$ system. $E_{\protect{K^-}p}$ is the
two-body energy of $\protect{K^-}p$.}
\label{tab4}%
\begin{tabular}{cccccc}
\hline\hline
$\bar{K}N$ & $a,$ fm & $\left\langle r^{2}\right\rangle ^{1/2}$, fm
& $E_{{K^-}p},$ MeV & $NN$ & $B,$ MeV \\ \hline
AY & 1.88 & 1.35 & 30.26 & AV14 & 47.3 \\ 
&  &  &  & MT & 46.0 \\ 
&  &  &  & T & 46.3 \\ \hline
HW & 2.68 & 1.94 & 11.16 & AV14 & 21.6 \\ 
&  &  &  & MT & 20.4 \\ 
&  &  &  & T & 20.6 \\ \hline\hline
\end{tabular}
\end{center} 
\end{table}
The analysis of the calculations presented in Table \ref{tab4} shows that
the AY potential as the $\bar{K}N$ interaction input fall into the
46 - 47 MeV range for\emph{\ }the binding energy\emph{\ }of the ${K^-}pp$ cluster, while the chiral HW $\bar{K}\emph{N}$ potential
gives about 20.4 - 21.6 MeV for the binding energy.\emph{\ }Now one can
address the theoretical discrepancies in the binding energy for
the ${K^-}pp$ system related to the different $NN$ and $KN$
interactions. The comparison of the results of calculations presented in
Tables \ref{tab2} and \ref{tab4} for the binding energy for the $
{K^-}pp$ system obtained by both methods are in reasonable agreement. The
ground state energy is not sensitive to the $NN$ interaction. However, it
shows very strong dependence on the kaon-nucleon potential. The energy of
the ground state, as well as the width calculated for the energy-independent 
${\bar K}N$ interaction \cite{AYPRC2002} are more than twice bigger than for the
energy-dependent chiral $\bar{K}N$ potential \cite{HW}. Therefore,
the highest binding energies are those that are obtained based on the
phenomenological AY potential. Discrepancies obtained for the binding energy
using the same potentials but different methods, - the method of HH in
momentum representation and the method of Faddeev equations in configuration
space,- are mostly related to a problem of an equivalent representation of
the potentials in momentum and configuration spaces.

Now let us compare our results with those obtained using variational methods,
the method of Faddeev equations in the momentum representation and the
method of hyperspherical functions in configuration space. The summary of
the results for the binding energy and the width with different theoretical
approaches and models, obtained for different kinds of $NN$ and $\bar
{K}N$ interactions, are presented in Table \ref{tab5}. 
\begin{table}[t]
\begin{center} 
\caption{Summary of the theoretical studies for the $\protect\bar{K}%
pp$ cluster.}
\label{tab5}%
\begin{tabular}{ccccc}
\hline\hline \\
Method & $B(\protect{K^-}pp)$ & Width, $\Gamma$ & $\protect\bar{K}N$ & References \\ 
& MeV & MeV &  &  \\ \hline
Variational & 48 & 61 & AY & \cite{AYPRC2002}, \cite{AYPL2002}, \cite{AYKN}
\\ 
Methods & 20$\pm3$ & 40-70 & Chiral model & \cite{DHWnp2008},\cite{DHW} \\ 
& 40-80 & 40-85 & Separable & \cite{WycechGreen} \\ 
& 20-35 & 20-65 & Chiral model & \cite{FeshbaxRes2014} \\ 
& 124 & 12 & AY & \cite{ChinesePhys} \\ \hline
Methods of & 47-70 & 90-100 & Separable En. Indep. & \cite{Nina1}, \cite%
{Nina2},\cite{Nina2014} \\ 
Faddeev & $\sim$32 & $\sim$50-65 & Separable En. Dep. & \cite{Nina2014} \\ 
equations & 45-95 & 45-80 & Separable En. Indep. & \cite{IkedaSato1}, \cite%
{IkedaSato2},\cite{IkedaKamanoSato3} \\ 
& 9-16 & 34-40 & Separable En. Dep. & \cite{IkedaKamanoSato3} \\ 
& 30-40 & 50-80 &  & \cite{FaddFixCent1} - \cite{FaddFixCent4} \\ 
& $\sim$52 &  & Separable En. Indep. & \cite{FadYa4body} \\ \hline
Methods of HH & $\sim$16 & $\sim$41 & Chiral model & \cite{BarneaK} \\ 
& 15-17 & 36-43 & Chiral model & \cite{RKSh.Ts2014} \\ 
& 40-48 & 75-96 & AY & \cite{RKSh.Ts2014} \\ \hline\hline
\end{tabular}
\end{center} 
\end{table}
We start the comparison with variational calculations. In Ref. \cite{AYKN},
the variational calculations were carried out by constructing a wave
function for the three-body system with correlation functions for each of
the constituent pairs on the basis of multiple scattering theory. The
binding energy and the width for the ${K}^-pp$ cluster was
calculated by employing the AY potential as the $\bar{K}N$
interaction and the bare Tamagaki G3RS potential \cite{Tpotential} as the $%
NN $ interaction. Thus, the authors of \cite{AYKN} used the same $s$-wave
interactions, and this allows us to compare our results with the previous
one. The binding energy found with the DFE and HH methods are in good
agreement with the one obtained with the variational method. This is a good
sign that the binding energy does not depend significantly on the method of
calculation. The prediction in Ref. \cite{WycechGreen} for the binding
energies of $I_{tot}$ = 1/2, $I_{NN}$ = 1 states given by the $s$-wave
interactions and described by multiple scattering in the single $\bar%
{K}N$ channel falls into the 40 - 80 MeV range with the parametrization for
the Argonne AV18 potential from Ref. \cite{ArgonneV18}. These results are
consistent with our calculations using the AV14
potential \cite{ArgonneV14} and the results from \cite{AYKN}. The
differences within this range are mostly due to a different $\bar{K}%
N $ input and possibly slightly due to the $NN$ input.

Different variational approaches used in Refs. \cite{AYKN} and \cite{DHW}
are of comparable quality in their high degree of consistency. Dot\.{e},
Hyodo and Weise \cite{DHW} employed several versions of energy-dependent
effective $\bar{K}N$ interactions derived from chiral SU(3) dynamics
together with the realistic Argonne 18 $NN$ potential \cite{ArgonneV18}.
They found that the antikaonic dibaryon ${K^-}pp$ is not deeply
bound and obtained 20$\pm 3$~Mev for the binding energy. Our calculations in
the framework of the DFE and HH methods when we employ the effective
energy-dependent chiral-theory based HW potential for $\bar{K}N$
interaction and different $NN$ interactions, as inputs, also predict a
shallowly bound ${K^-}pp$ cluster. This is consistent with results
from \cite{HW}, \cite{DHW} and recent calculations \cite{FeshbaxRes2014}.

The first calculations with Faddeev equations in Alt-Grassberger-Sandhas
form \cite{AGS} for the three-body system with coupled $\bar{K}NN$
and $\pi \Sigma N$ channels were performed in Refs. \cite{Nina1}, \cite%
{Nina2} for the study of the $\bar{K}NN$ system with separable
two-body potentials yield large bindings. In Refs. \cite{Nina1, Nina2} authors obtained 55--70 MeV, and 95--110 MeV for the binding energy
and the width, respectively. With a similar approach, Ikeda and Sato \cite%
{IkedaSato1} calculated $B\thicksim $80 Mev and $\Gamma \thicksim $73 MeV.
Later, two of the authors of \cite{IkedaSato1} repeated their calculation in 
\cite{IkedaSato2, IkedaKamanoSato3} using two models with the
energy-independent and energy-dependent potentials for the $s-$wave $\overset%
{\_}{K}N$ interaction, and their calculations yield smaller values for the
binding energy 44-58 MeV and width 34-40 MeV \cite{IkedaKamanoSato3}.
Recently in Ref. \cite{Nina2014}, the Faddeev calculations for the $\overset{%
\_}{K}NN$ quasi-bound state with the two phenomenological and the
energy-dependent chirally motivated models of the $\bar{K}N$
interaction were carried out. The binding energy for the ${K^-}pp$
cluster obtained was 32 MeV with the chirally motivated models and 47 - 54
MeV with the phenomenological $\bar{K}N$ potentials. The width is
about 50 - 65 MeV. Therefore, we can conclude that the Faddeev calculations
for the energy-independent models for the $\bar{K}N$ interaction 
\cite{Nina1, Nina2, IkedaSato2, IkedaKamanoSato3}
predict a deeper binding energy than that of the energy-dependent
description of the $\bar{K}N$ interaction \cite{IkedaKamanoSato3, Nina2014}. Our calculations obtained with both methods confirm that
the effective $\bar{K}N$ interaction derived from chiral SU(3)
dynamics yields a shallowly bound ${K^-}pp$ cluster, while the
phenomenological energy independent AY potential predicts much deeper
binding energy for all considered $NN$ interactions. A more detailed
comparison of our results with the Faddeev calculations \cite{Nina1, Nina2, IkedaSato1, IkedaSato2, IkedaKamanoSato3, Nina2014} is not easy because these calculations use separable potentials
to describe the $\bar{K}N$ and $NN$ interactions.

The fixed center approximation (FCA)\ to the Faddeev equations \cite%
{FaddFixCent1, FaddFixCent2, FaddFixCent3, FaddFixCent4} leads to the binding energy of the $%
{K^-}pp$ system with $S=0$ 30-40 MeV and the width 50-80 MeV that
includes the recent improvement of the method \cite{FaddFixCent3, FaddFixCent4} by including the charge exchange mechanisms in the ${K^-}$ rescattering and absorption which have been ignored in previous
works within this approximation. Due to the absorption of ${K^-}$
by two nucleons, the width of the bound ${K}^-pp$ cluster is
increased by about 30 MeV \cite{FaddFixCent4}. Although the FCA to the
Faddeev equation makes a static picture of the two nucleon and does not
consider the recoil of the spectator nucleon, this approach provides
reasonable values for the binding energy and width, as seen from Table \ref%
{tab5}, and qualitatively corroborates findings done with other methods
which are technically much more involved.

The study of ${K^-}pp$ with $S=0$ was performed using the method
of hyperspherical functions in configuration space \cite{BarneaK} and in the
momentum representation \cite{RKSh.Ts2014}. In Ref. \cite{BarneaK}, the
binding energies and widths of the three-body ${K}^-pp$ cluster
are calculated using the realistic AV4 \cite{AV4} $NN$ potential and a
subthreshold energy-dependent chiral $\bar{K}N$ interaction derived
with a chiral model \cite{HW}. The results are in good agreement with
previous ${K^-}pp$ calculations \cite{DHWnp2008, DHW} with
an energy-dependent chiral $\bar{K}N$ interaction as input.
Calculations of the binding energy and the width with the method of HH in
the momentum representation \cite{RKSh.Ts2014} reproduced the results of 
Ref. \cite{BarneaK}, and we assume the slight difference arised from the
difference between using AV4 and AV18 potentials and perhaps is related to
the conversion of the hyperspherical expansion. However, most importantly,
our results and those of the previous HH calculations \cite{BarneaK, RKSh.Ts2014} are in 
good agreement and support the conclusion that the key
role in binding the ${K^-}pp$ system is played by the $\bar%
{K}N$ interaction and the $\bar{K}N$ potential obtained based on
chiral SU(3) dynamics leads to binding energies of relatively low values.
Also, these calculations show that the binding energies are small and the
widths are more than twice that of the binding energies. Therefore, all
model calculations with the $\bar{K}N$ interaction derived based on
the chiral unitary approach predict a shallow binding state with very large
width for the ${K^-}pp$ cluster. This is a precautionary
indication that it may be difficult to experimentally observe the ${K^-}pp$ cluster.

\section*{Conclusion}

\bigskip Within the framework of a potential model for the kaonic cluster $
{K^-}pp,$ we perform nonrelativistic three-body calculations using
two methods: the method of hyperspherical harmonics in
the momentum representation and the method of Faddeev equations in
configuration space. We examine how the quasi-bound state of the ${K^-}pp$ cluster depends on different choices of the $\bar{K}N$ and $%
NN$\ interactions. Our consideration includes the realistic AV14 \cite%
{ArgonneV14}, the semi-realistic Malfliet and Tjon MT-I-III \cite{MT} and
Tamagaki G3RS \cite{Tpotential} potentials as input for the $NN$ interaction
and we employ the phenomenological AY potential \cite{AYPRC2002, AYKN}
and HW potential \cite{HW} based on chiral SU(3) dynamics as input for the $%
\bar{K}N$ interaction. The results for the binding energy of the $%
{K^-}pp$ system obtained by the method of HH and the DFE method
are in reasonable agreement. A discrepancy obtained is related to a problem
of the equivalent representation of the potentials in momentum and
configuration spaces. For all types of considered $\ NN$ interactions, both
methods predict deeply bound states for the AY $\bar{K}N$
interaction and a relatively shallowly bound ${K^-}pp$ cluster for
the effective $\bar{K}N$ interactions derived from chiral SU(3)
dynamics. Moreover, the ${K^-}pp$ cluster is the most strongly
quasi-bound three-body system. The results of our calculations within the
DFE and HH methods show that the binding energy of the ${K^-}pp$
system depends entirely on the ansatz for the $\bar{K}N$ interaction
and substantially changes when we use a phenomenologically constructed $%
\bar{K}N$ potential \cite{AYKN} and a $\bar{K}N$ potential
obtained within the framework of the chiral unitary approach \cite{HW, Weise}. Perhaps the ambiguity of the $\bar{K}N$ interaction stems
from an accuracy of description of experimental data for the energy shift in
the kaonic hydrogen atom and ${K^-}p$ scattering by the $\bar{K}N$ potentials. Related to the sensitivity of the binding energy to the
details of the $NN$ potentials Ref. \cite{HW} \ mentioned that as long as
the ${K^-}pp$ system is only weakly bound, the dependence on
different types of $NN$ interactions is weak. In fact, our study confirms
this conclusion and, moreover, shows that the dependence on different types
of $NN$ interactions is also weak if the ${K^-}pp$ system is
strongly bound. Using the formalism of the Faddeev equations, we may separate 
channel configurations of the total wave function of the system. We have shown 
that the configuration (${K^-}p$) + $p$ with the singlet isospin state of the pair ${K^-}p$ 
dominates in the system due to strong  ${K^-}p$ interaction. We have seen that, on the 
background of the $\bar{K}N$ potential, different $NN$ interactions weakly effluent to 
the bound state energy. It has to be noted that the $s$-wave models, 
with the simplifications taken, provide reasonable description of the system.
The analysis of data presented in Table \ref{tab5} shows
that the width is always larger than the binding \ energy. Particularly for
some calculations, the width is more than twice as much as the binding
energy. Thus, we are facing a situation in which the states have a much
larger width than binding energy, which makes the experimental observation
challenging \cite{FaddFixCent4} and it may be hard to identify the
resonance. However, the continuation of the experimental search for the
quasi-bound kaonic cluster still remains important.

\section*{Acknowledgements}
This work is supported by the NSF (HRD-1345219) and NASA (NNX09AV07A). R.Ya.K is partially supported by MES RK, the grant 3106/GF4. The
numerical calculations were performed at the High Performance Computing Center of the North Carolina State University and the Center for Theoretical Physics of New York City College of Technology, CUNY.

%\section*{References}

\end{document}